# Scaling and Universality in Models of Step Bunching: The "$C^+ - C^-$" Model

V. Tonchev[1*], B. Ranguelov[1*], H. Omi[2], A. Pimpinelli[3]

[1]Institute of Physical Chemistry, Bulgarian Academy of Sciences, 1113 Sofia, Bulgaria
[2]NTT Basic Research Laboratories, NTT Corporation, Atsugi, Kanagawa 243-0198, Japan
[3]LASMEA, CNRS/Université Blaise-Pascal, Aubiére, France

**Abstract**
We recently introduced a novel model of step flow crystal growth – the so-called "$C^+ - C^-$" model [B. Ranguelov *et al.*, CR Acad. Bul. Sci. **60**, 389 (2007)]. In this paper we aim to develop a complete picture of the model's behaviour in the framework of the notion of universality classes. The basic assumption of the model is that the reference ("equilibrium") densities used to compute the supersaturation might be different on either side of a step, so $C_L/C_R \neq 1$ ($L/R$ stands for left/right in a step train descending from left to right), and that this will eventually cause destabilization of the regular step train. Linear stability analysis considering perturbation of the whole step train shows that the vicinal is always unstable when the condition $C_L/C_R > 1$ is fulfilled. Numerical integration of the equations of step motion combined with an original monitoring scheme(s) results in obtaining the exact size- and time- scaling of the step bunches in the limit of long times (including the numerical prefactors). Over a broad range of parameters the surface morphology is characterized by the appearance of the minimal interstep distance at the beginning of the bunches (at the trailing edge of the bunch) and may be described by a single universality class, different from those already generated by continuum theories [Pimpinelli *et al.*, PRL **88**, 206103 (2002), Krug *et al.*, PRB **71**, 045412 (2005)]. In particular, the scaling of the minimal interstep distance $l_{min}$ in the new universality class is shown to be $l_{min} = (S_n/N)^{1/(n+1)}$, where $N$ is the number of steps in the bunch, $n$ is the exponent in the step-step repulsion law $U \sim 1/d_0^n$ for two steps placed a distance $d_0$ apart and $S_n$ is a combination of the model parameters. It is also shown that $N$ scales with time with universal exponent $1/2$ independent of $n$. For the regime of slow diffusion it is obtained for the first time that the time scaling depends only on the destabilization parameter $C_L/C_R$. The bunching outside the parameter region where the above scaling exists cannot be assigned to a specific universality class and thus should be considered non-universal.

## I. Introduction

Monatomic steps on crystal surfaces appear when the crystal is cut along a plane that is not parallel to the atomic planes. These steps are the subject of longstanding interest both from a technological and a fundamental point of view. So-called *step flow growth*, in which the crystal grows via attachment of atoms to the existing network of steps and **not** through nucleation of islands on the terraces between the steps - is the growth mode of technological importance. This is also the reason for intensive theoretical work involving playing with steps and their movement. One independent direction is the study of unstable step flow growth leading to two general instabilities – step bunching, when the initial equidistant step spacing is lost and steps group to form *bunches*, and step meandering, when the initially straight steps bend to form *meanders* similar to those formed by rivers in nature.

The interest in introducing novel models of surface instabilities has recently been stimulated further by experiments on step flow growth of Cu- [1,2] and Si- vicinals [3] in which simultaneous step bunching and step meandering was observed. This phenomenon is quite unexpected and contradicts the contemporary paradigm of surface instabilities. The reason is that according to the present concepts, the normal Schwoebel effect in growth, i.e. adatoms attach to the steps preferentially from the lower terrace, results in step meandering, while the inverse Schwoebel effect, in which adatoms attach to the steps preferentially from the higher terraces, results in step bunching. In principal, besides the destabilization from various sources, all of the models of step bunching instability contain also a stabilizing factor due to the step-step repulsion. Typically the destabilizing factors are the electromigration force acting on adatoms or the Ehrlich-Schwoebel effect. Step-step repulsions of different nature stabilize and promote the equidistant step distribution. In the models, the step-step repulsion affects the equilibrium concentrations used to compute the supersaturation/undersaturation and thus, the step velocity in growth/sublimation. Here we present a new model in which a difference of the "equilibrium" adatom concentrations on both sides of the step is assumed. In this manner the destabilization and stabilization are "intrinsic" and a consequence of the same source – the step-step repulsion. In the new model the shape of the bunches formed as a result of the destabilization is also different – the minimal interstep distance appears in the beginning (trailing edge) of the bunch while in the bunches formed by 'classical' mechanisms it is found in the middle of the bunch. Quantitatively, an important characteristic of the new type of step bunching is that the exponent $\gamma$ of the average interstep distance $l_b$ in the size-scaling relation $l_b \sim N^{-\gamma}$, $N$ being the number of steps in the bunch, is $\gamma = 0.29\pm0.05$ [1,2] while in most cases of experimental and theoretical studies it was found to be 2/3 [4,5,6,7] or at least larger than 1/2 in the case of bunching of the so called "transparent steps" [8,5].

Here we study the linear stability of the whole step train using the equations of step motion obtained within the "$C^+ - C^-$" model [9]. Then, we integrate these equations numerically applying an original monitoring scheme to obtain the size- and time-scaling relations for the step bunches formed in the long-time limit.

Our results represent a challenge for further theoretical and experimental efforts to clarify and extend the existing classification scheme of universality classes in bunching proposed by Pimpinelli-Tonchev-Videcoq-Vladimirova (PTVV) [10,7].

## II. Model

Following the pioneering work [11] of Burton, Cabrera and Frank (BCF) various modifications of their model of vicinal growth have been made in order to calculate more adequately the step velocity as a function of the actual supersaturation. Now, it is common to calculate the actual adatom density in the vicinity of the step and then to compare it to a reference density which is in general the equilibrium one. The basic assumption of the "$C^+ - C^-$" model is that the "reference" densities close to the step edges are not the equilibrium densities, and in fact are different on either side of the step. This idea can be traced back to the works of Suga et al. [12] and O. Pierre-Louis [13]. However, these authors do not follow the consequences of this assumption on the long-time behavior of the step system. Such a difference in the "equilibrium" concentrations on both sides of the step should always be expected in an experimental context when the surface structure is different in the vicinity of the step on both terraces [14]. In what follows we will always write "reference density" and not "equilibrium" density to avoid misunderstanding – in thermodynamic equilibrium there can exist only one equilibrium density but in non-equilibrium conditions this restriction may be avoided. The model was also obtained [23] as an effective step-flow model from a two-particle model [15,16]. The two-particle model have been shown [15] by kinetic Monte Carlo simulation to produce in a given range of model parameters simultaneous step bunching and step meandering.

Difference in the "equilibrium" concentrations may appear also in more complex contexts, where other destabilizing mechanisms like adatom electromigration, Ehrlich-Shwoebel effect, etc., could also come into play. In this study we follow the consequences of only one, new destabilizing mechanism and obtain the new static and dynamic scaling laws for this "clear" situation. Thus one can judge, when dealing with complex environments, which mechanism prevails. For example, the size scaling of the minimal interstep distance is $l_{min} \sim (1/N)^{2/(n+1)}$ for a class of step bunching models [7] while for the new model it is $l_{min} \sim (1/N)^{1/(n+1)}$, where $n$ is the power in the step-step repulsions law $U \sim 1/d_0^n$, $d_0$ being the interstep distance.

The "$C^+ - C^-$" model is a 1D extended BCF model with steps descending in the positive $x$ direction. The density of the adatoms $C(x)$ on an isolated terrace is obtained assuming that the concentration fields on neighboring terraces are coupled only through the step acting as a source/sink of adatoms. The additional assumption that the adatom density adjusts instantaneously to the step positions allows us to consider the steps as motionless. The stationary diffusion equation for this case is rather simple:

$$D_s \frac{\partial^2 C}{\partial x^2} + F = 0, \tag{1}$$

where $F$ is the flux towards the crystal surface and $D_S$ is the surface diffusion coefficient of the adatoms while the desorption is neglected. Equation (1) is directly integrated and the two integration constants are obtained using boundary conditions on the two steps bordering the terrace, placed at $x_i$ and $x_{i+1}$ respectively:

$$D_s \frac{\partial C}{\partial x}\bigg|_{x=x_i} = K[C(x_i) - C_e^+(x_i)] \quad \text{and} \quad D_s \frac{\partial C}{\partial x}\bigg|_{x=x_{i+1}} = -K[C(x_{i+1}) - C_e^-(x_{i+1})], \tag{2}$$

where $K$ is the kinetic coefficient for adatom attachment/detachment to/from the step, that we assume to be equal on both sides of the step. The reference densities $C_L$ and $C_R$ are affected by step-step repulsion with energy $U = A/d_0^n$ ($A$ is the magnitude of the repulsion and $d_0$ is the interstep distance) and are given by:

$$C_e^+(x_i) = C_R \left[1 + l_0^{n+1}\left(\frac{1}{(x_{i+1}-x_i)^{n+1}} - \frac{1}{(x_i-x_{i-1})^{n+1}}\right)\right] \tag{3}$$

$$C_e^-(x_{i+1}) = C_L \left[1 + l_0^{n+1}\left(\frac{1}{(x_{i+2}-x_{i+1})^{n+1}} - \frac{1}{(x_{i+1}-x_i)^{n+1}}\right)\right] \tag{4}$$

with the characteristic length of the interstep repulsion $l_0 = (nA\Omega/kT)^{\frac{1}{n+1}}$, where $\Omega$ is the crystal surface area per atom. Usually $n$ takes the *canonical* value [17] $n = 2$ but leaving $n$ free to vary allows us to study systematically the scaling behavior of the model. The basic assumption of the model is that the reference densities on both sides of the step, $C_L$ and $C_R$, are not equal to the equilibrium density, and are not determined self-consistently. Instead, we expect them to be fixed by the coupling to other diffusion fields, such as diffusing species, or by an external condition, such as a step region with different structure, and different diffusion or binding energetic, from the rest of the surface. We will then write:

$$C_L \neq C_R \tag{5}$$

This condition will be shown to eventually lead to instability of the growing surface. The concentration profile on the terrace is obtained as:

$$C(x) = \frac{F}{2D_s}[(x_{i+1}-x)(x-x_i)+d(x_{i+1}-x_i)]+ \tag{6}$$
$$+\frac{(x-x_i)[C_e^-(x_{i+1})-C_e^+(x_i)]+d[C_e^-(x_{i+1})+C_e^+(x_i)]+C_e^+(x_i)(x_{i+1}-x_i)}{2d+x_{i+1}-x_i}$$

where $d \equiv D_s/K$ is the so called *kinetic length* [18]. Furthermore, one obtains the step velocity as a sum of two contributions – the mass fluxes from the two terraces neighboring a step:

$$\frac{1}{D_s\Omega}\frac{dx_i}{dt} = \frac{\partial C^i(x_i)}{\partial x} - \frac{\partial C^{i-1}(x_i)}{\partial x} \tag{7}$$

which finally yields:

$$V_i = \frac{dx_i}{dt} = \Omega D_s\left(-\frac{\partial C}{\partial x}\bigg|^{i-1}_{x=x_i} + \frac{\partial C}{\partial x}\bigg|^{i}_{x=x_i}\right) = \Omega D_s\left(C_1^i - C_1^{i-1}\right) = \tag{8}$$
$$F\Omega\left(\frac{x_{i+1}-x_{i-1}}{2}\right)+\frac{\Omega D_s}{2d+\Delta x_{i+1}}\left\{C_L\left[1+l_0^{n+1}\left(\frac{1}{\Delta x_{i+2}^{n+1}}-\frac{1}{\Delta x_{i+1}^{n+1}}\right)\right]-C_R\left[1+l_0^{n+1}\left(\frac{1}{\Delta x_{i+1}^{n+1}}-\frac{1}{\Delta x_i^{n+1}}\right)\right]\right\}-$$
$$-\frac{\Omega D_s}{2d+\Delta x_i}\left\{C_L\left[1+l_0^{n+1}\left(\frac{1}{\Delta x_{i+1}^{n+1}}-\frac{1}{\Delta x_i^{n+1}}\right)\right]-C_R\left[1+l_0^{n+1}\left(\frac{1}{\Delta x_i^{n+1}}-\frac{1}{\Delta x_{i-1}^{n+1}}\right)\right]\right\}$$

So, the velocity of the *i*-th step is obtained as a function of the model parameters: the flux $F$, surface diffusion constant $D_s$, the step kinetic coefficient $K$ magnitude of the interstep repulsions $A$. The differences of type $\Delta x_i \equiv x_i - x_{i-1}$ are the widths of the corresponding terraces (here, the terrace between the *i-th* and the *i-1-th* step). Equation (8) is transformed into a dimensionless form containing four dimensionless parameters:

$$\frac{d}{l} = \frac{1}{l}\frac{D_s}{K};\ \frac{C_L}{C_R};\ \frac{D_sC_R}{Fl^2};\ \frac{l_0}{l} = \frac{1}{l}\left(\frac{nA\Omega}{kT}\right)^{\frac{1}{n+1}},\text{ where } l \text{ is the initial vicinal distance.}$$

### III. Results

The set of equations (8) dictates the dynamical behavior of the step system. A solution can always be found, describing equidistant steps travelling at constant velocity. The stability of this solution has to be investigated, and two situations may arise. The solution is stable - any perturbation in the terrace widths will die away - or unstable - in which case, the vicinal surface with equal distances between the steps decomposes into regions with high step density (called "bunches") and regions with few or no steps (called "terraces").

### III.1. Linear Stability Analysis.

The simplest version of linear stability analysis was carried out [9] by perturbing only the *i*-th step in the moving equidistant step train and thus changing the widths of the two neighboring terraces:

$$\Delta x_i \to l + \delta l, \Delta x_{i+1} \to l - \delta l$$

Then an instability condition was obtained from the equations for step velocity, Eq.(8), in the form:

$$\left(\frac{2}{3(n+1)}\right) > \left(\frac{l_0}{l}\right)^{n+1} \left(\frac{2d/l+1}{(C_L - C_R)/(C_L + C_R)}\right) \equiv S_n \quad (9)$$

Thus, the linear stability of a terrace width in the regular step train is independent of the ratio $D_s C_R / Fl^2$. Further in this paper we will show that this parameter also does not enter into the scaling relations describing the late stages of the bunching process and that what determines the system's behavior in this stage is only the combination of parameters denoted by $S_n$.

A more general procedure consists [19] of perturbation of the whole step train in the form $\delta \Delta x_i = e^{jiq} x_q$, which formally is a perturbation of the terrace widths. After performing the algebraic operations, we obtain the usual expression for the time evolution of the perturbation $x_q$ with real part

$$dx_q(t)/dt = \left(B_2 q^2 - B_4 q^4\right) x_q(t) \quad (10)$$

where:

$$B_2 = \frac{\Omega D_S (C_L - C_R)}{(2d+l)^2}; B_4 = \frac{\Omega D_S (C_L - C_R) l^{n+2} + 6\widetilde{A}\Omega D_S (n+1)(C_L + C_R)(2d+l)}{12(2d+l)^2 l^{n+2}}$$

Now, what determines the instability of the step train is the condition $B_2 > 0$ and this is always fulfilled if $C_L > C_R$. The perturbations with wave numbers $q < q_{\max}$ will grow further, where:

$$q_{\max} = \left[\frac{12}{1 + 6(n+1)S_n}\right]^{1/2}. \quad (11)$$

Thus we see that the analysis based on the linear stability of a single terrace width is modified, though it contains the same scaling parameter $S_n$.

## III.2. Numerical Calculations

We integrate numerically the equations of step motion using a fourth order Runge-Kutta procedure [20] and obtain step positions at consecutive moments in time. These step trajectories are shown in Fig. 1. Two interesting features are also presented on this figure: just before the coalescence of two bunches, the front one temporarily moves in the direction opposite to the growth – denoted by the two arrows. Also, the regular exchange of monatomic steps between bunches starts after some time – a phenomenon originally observed [21] in a different model [22].

The area denoted by the larger arrow in Fig. 1 (around step position = 100) is presented by means of the surface slope in Fig. 2. It is clearly seen that the surface slope is largest at the beginning of the bunches (in their trailing edge) where individual steps join the bunch after leaving the one from behind and crossing the terrace. Thus the minimal interstep distance is the first one in the bunch and this behaviour is preserved during the entire growth process.

Prior to the explanation of the monitoring schemes, an important issue from the core of the calculations should be addressed - how to decide whether part of the simulated vicinal surface is within a bunch or part of the terrace (see also Figure 3b where in particular a terrace between two bunches is drawn). We choose a rather natural definition – if a distance between two steps is smaller than the initial (vicinal) one this is a *bunch distance*.

In order to collect quantitative information on the bunching process we adopt two simultaneous monitoring schemes illustrated in Figure 3a.
**Monitoring Scheme I (MS-I):** ("*snapshot monitoring*") – the average bunch size, average bunch width, average interstep distance in the bunch, minimal bunch distance – the minimal interstep distance in the entire step system, the average terrace width and the average distance between two steps on a terrace are computed at fixed time intervals. To be more concrete, our computer code does the following – given the number of steps, and hence the number of distances, it checks how many of them are less than the initial (*vicinal*) distance and then counts the number of bunches. Starting with the first interstep distance in the simulation box, all other distances are visited and when the first distance that is less than the vicinal one occurs, this is the beginning of the first bunch. When two adjacent distances are less then the initial one, they belong to the same bunch. This bunch ends when a distance larger than the initial (vicinal) one is encountered. If two such distances are adjacent, they belong to the same terrace. This monitoring scheme is used in experiments to study the time behavior of the average terrace width.

**Monitoring Scheme II (MS-II):** ("*cumulative monitoring*") – for any bunch size that may appear during the whole integration time the code cumulates the following quantities: bunch width, minimum bunch distance and the first and last bunch distances with respect to the direction of step movement. This is done in the following manner – every time a bunch with given size $N$ is encountered, all quantities are updated for this size $N$ and the new average values are stored. This monitoring scheme is the same as the one used to study bunching experimentally [5].

In Figure 3b the bunch quantities that are to be monitored are shown. The initial vicinal angle is $\phi$ and it determines the initial vicinal distance $l$. Individual monatomic steps have height $h_0$. Bunch size $N$ and bunch width $L_b$ are shown. The smallest distance in the bunch (which in this model appears to be the first one) is $l_{min}$.

An example of the results obtained from the two monitoring schemes is given in Fig. 4 and the strict equivalence is clearly seen. This is comforting, since only the MS-I scheme allows us to study both the time- and size- scaling of the step bunches.

**III.3 Scaling Relations for the Step Bunches**

In the beginning of this Section we start with the results from MS-II. The dependence of the minimal bunch distance $l_{min}$ on the number of steps $N$ in the bunch is shown in Fig. 5 for different values of $n$, the exponent in the step-step repulsion law. The obtained dependence $l_{min} \sim N^{-1/(n+1)}$ differs from that of the previously studied case of bunching due to electromigration or the Ehrlich-Schwoebel effect [7] which is $l_{min} \sim N^{-2/(n+1)}$. In the latter case the minimum interstep distance appears in the middle of the bunch and the scaling behavior of the *first* bunch distance is $l_1 \sim N^{-1/(n+1)}$.

In other words, if one looks at the scaling of the interstep distance at the trailing edge of a bunch, all models yield the same result, $l_1 \sim N^{-1/(n+1)}$. However, for the "$C^+ - C^-$" model $l_1$ is also the minimum interstep distance. On Fig. 6 are plotted the data for $l_{min}$ for different values of $d/l$ ranging from 0.001 to 100.

Instead of the bunch size, we put on the x axis the rescaled bunch size – the quantity $(S_n/N)^{1/(n+1)}$ (also shown in Fig. 5 as dashed lines that fit the data). It is seen that for bigger sizes the data collapse onto the $y = x$ line, so that $(S_n/N)^{1/(n+1)}$ is the correct scaling parameter. It should be noted that in the latter scaling form of $l_{min}$ there is no numerical pre-factor.

As shown in Fig. 7 the data for $C_L/C_R < 2$ significantly deviate from the obtained scaling, thus, in fact, leaving the universality class still at the stage of size-scaling studies. Our attempt to find some systematic description of the size scaling for such small values of $C_L/C_R$ failed. Thus, we consider it non-universal. After finding the exact scaling for $l_{min}$, we also analyzed the size scaling of the bunch width $L_b$ and we found that its form is:

$$L_b/l = \frac{(n+1)}{n}\left(S_n N^n\right)^{\frac{1}{n+1}} \qquad (12)$$

Thus, the ratio between the average and the minimal bunch distances is obtained as:

$$\frac{L_b/N}{l_{min}} = \frac{l_b}{l_{min}} = \frac{n+1}{n} \qquad (13)$$

Further, in order to find the time-scaling for the step bunches and thus to complete the characterization of the universality class, we plot the results obtained using MS-I. One particular result is shown in Figure 8, for specific values of the parameters.

The most important conclusion for the time dependencies is that the time scaling exponent of the bunch size $N$ is always $1/2$, independent of $n$. Thus the study of the model universality is completed and compared with the existing paradigm in Table1. Finally we are able to determine the exact form of the time dependence of the bunch size that contains only the destabilization parameter $C_L/C_R$. Figure 9 shows the time scaling of the bunch size $N$ for $C_L/C_R \gg 1$ and $d/l \ll 1$, given as $N = \sqrt{3(C_L/C_R)t}$. For lower values of $C_L/C_R$, the pre-factor decreases and for example when $C_L/C_R = 10$ it is 2.6 ($N = \sqrt{2.6(C_L/C_R)t}$). The exact dependence of the numerical pre-factor on $C_L/C_R$ for values of this parameter lower than ~20 remains to be studied. The increasing of the kinetic length $d/l$ preserves the exponent $1/2$ but modifies significantly the intercept of this dependence in a manner that must be studied in more detail.

Similar time scaling of the bunch size $N$, containing only the destabilization parameter and dependence of the numerical pre-factor on the destabilization parameter, was observed [24] for another model – that of Liu and Weeks [22] relevant for the case of step bunching due to electromigration in the kinetic limited regime.

## III. 4 Universality

A continuum equation for the time evolution of the surface height $h(x,t)$ has been proposed [10,7]:

$$\frac{\partial h}{\partial t} + \frac{\partial}{\partial x}\left[K_1 m_x^\rho - \frac{K_2}{m_x^k}\frac{\partial^2}{\partial x^2} m_x^n\right] = const \quad (14)$$

where $m \equiv \partial h(x,t)/\partial x$ is the surface slope, $K_1$, $K_2$ are constants that contain model parameters, $\rho$ is introduced as a free parameter in order to generalize the model and $k$ accounts for the two possible limiting regimes – $k = 0\,(\text{or}\,1)$ is supposed to be responsible for a diffusion (or attachment/detachment) limited regime. In Table 1 (third column) the resulting scaling relations in the PTVV universality picture are given for $k = 0$. In the second column of Table 1 are given the scaling relations obtained in the present study, valid for the whole range of studied parameters. It is a matter of future research to find in a systematic manner a suitable continuum equation that generates the new universality class.

## IV. Conclusion

The "$C^+ - C^-$" model is studied, expecting to recover the behaviour of a class of step bunching phenomena characterized qualitatively by the appearance of the minimal interstep distance at the beginning of the bunch instead of the usual appearance approximately in the middle of the bunch. Quantitatively, the size- and time-scaling exponents are different from those predicted by PTVV[10,7] and fall into a new universality class. Thus, the model is one of the reference models for further extension of the universality classes' picture of step bunching phenomena and challenges both theorists and experimentalists.

## V. Acknowledgements

Most calculations in this study were done on a computer cluster built with financial support of grants F-1413/2004, BM9/2006 and TK-X-1713/2007 from the Bulgarian National Science Fund and VIRT/NANOPHEN-FP6-INCO-CT-2005-016696. BR and VT acknowledge the support of MADARA (RNF01/0110) and IRC-CoSim Projects. VT acknowledges the hospitality and very stimulating working conditions at NTT Basic Research Laboratories, NTT Corporation, Morinosato Wakamiya, Japan.

24. V. Tonchev, B. Ranguelov, J. Krug, unpublished

Table 1. The new universality class is compared with those obtained from continuum theories.

| Scaling relation | "$C^+ - C^-$" model | PTVV with $\rho = -1$ | Universality classes PTVV [10] |
|---|---|---|---|
| $1/\alpha : L_b \sim N^{1/\alpha}$ | $n/(n+1)$ | $(n+1)/(n+3)$ | $(n-\rho)/(2+n-\rho)$ |
| $\beta = \alpha/z : N \sim t^\beta$ | $1/2$ | $1/2$ | $(2+n-\rho)/2(n+1-2\rho)$ |
| $1/z : L_b \sim t^{1/z}$ | $n/2(n+1)$ | $(n+1)/2(n+3)$ | $(n-\rho)/2(n+1-2\rho)$ |
| $\delta : l_b \sim t^\delta$ | $1/2(n+1)$ | $1/(n+3)$ | $1/(n+1-2\rho)$ |
| $\gamma : l_b \sim N^{-\gamma}$ | $1/(n+1)$ | $2/(n+3)$ | $2/(2+n-\rho)$ |

Table 1. Scaling relations for the new "$C^+ - C^-$" model, compared with those obtained from continuum theories. In the third column the number given corresponds to $\rho = -1$.

**Figure captions:**

Figure 1. Step trajectories. The arrows point to two areas where the front bunch temporarily moves in the direction opposite to the growth direction during the process of bunch coalescence. The transition from a surface without steps on the terraces towards surfaces with regular step exchange among bunches is also clearly seen. Model parameters: $d/l = 0.001$, $C_L/C_R = 10$, $l_0/l = 0.201$, $D_s C_R / Fl^2 = 1$. 100 of 1000 steps are shown. The slope of the surface shown with the longer arrow is shown separately in FIG 2.

Figure 2. Surface slope obtained for the "$C^+ - C^-$" model.

Figure 3a. Illustration of the two monitoring schemes used to study the bunching process. Four surface profiles are shown representing the time evolution of the vicinal surface. The monitoring scheme I (MS-I) collects information only from a given profile thus being a snapshot of the development of the instability, while monitoring scheme II (MS-II) covers the entire bunching process.

Figure 3b. Sketch of a bunched vicinal surface with the monitored quantities during the calculation.

Figure 4. Bunch Width $L_b$ vs. Bunch Size $N$ for different values of $n$ obtained from the two monitoring schemes. The fitting curves are drawn according to the scaling obtained later. Model parameters: $d/l = 0.001$; $C_L/C_R = 10$, $l_0/l = 0.214$, $D_s C_R / Fl^2 = 1$.

Figure 5. Minimal interstep distance in the bunch $l_{min}$ as a function of the bunch size $N$ (monitoring scheme II). For different values of $n$ the values of the step-step repulsion parameter $l_0$ are also different. The values of the scaling parameter $S_n$ are also shown. Model parameters: $d/l = 0.001$, $C_L/C_R = 10$, $l_0/l = 0.201$, $D_s C_R / Fl^2 = 1$.

Figure 6. The scaling relation for the minimal bunch distance $l_{min}$ is obtained, searching for the best collapse of the data for various values of $d/l$, $C_L/C_R = 10$, $l_0/l = 0.1$, $D_s C_R / Fl^2 = 1$ (only for $d/l = 100$, $D_s C_R / Fl^2 = 0.01$), monitoring scheme II.

Figure 7. The scaling relation for the minimal bunch distance $l_{min}$ is obtained, searching for the best collapse of the data for various values of the ratio $C_L/C_R$, $d/l = 0.1$, $l_0/l = 0.5$, $D_s C_R / Fl^2 = 1$, MS-II. It is seen that the data for small values $C_L/C_R$ significantly deviate from the obtained scaling, Fig 5.

Figure 8. Time dependencies of the bunch size $N$ and bunch width $L_b$. Model parameters: $d/l = 0.001$, $C_L/C_R = 10$, $l_0/l = 0.1$, $D_s C_R / Fl^2 = 1$.

Figure 9. Time dependence of bunch size N for different degrees of stabilization $C_L/C_R$. Other parameters of the model: $d/l = 0.001$, $l_0/l = 0.5$, $D_s C_R / Fl^2 = 1$, $n = 2,3$.

**Figures:**

Figure 1:

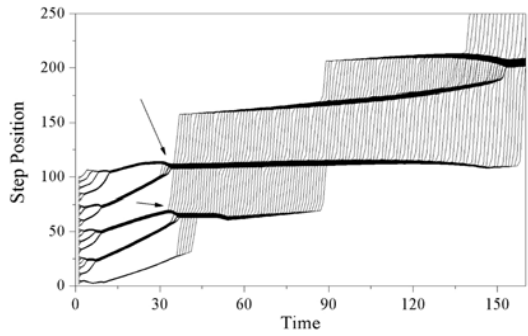

Figure 2:

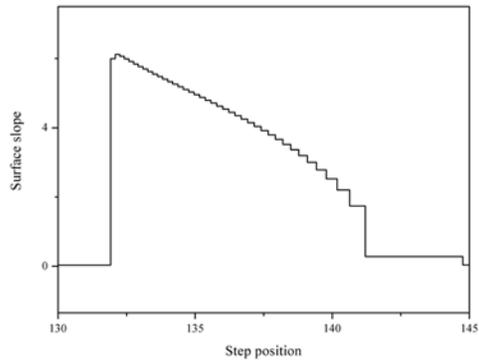

Figure 3a:

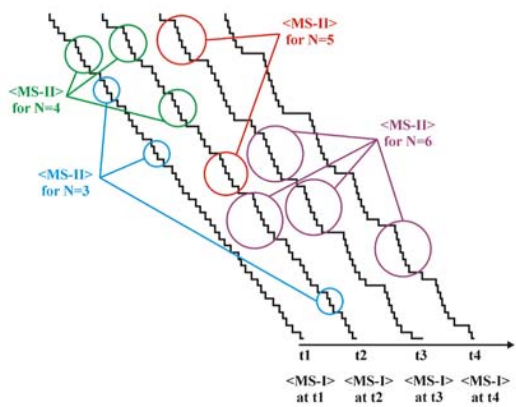

Figure 3b:

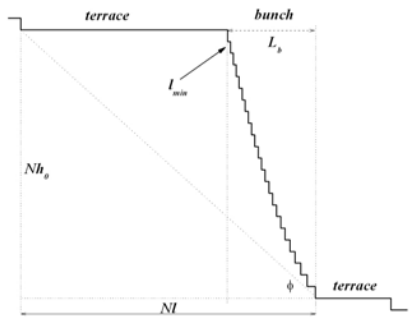

Figure 4:

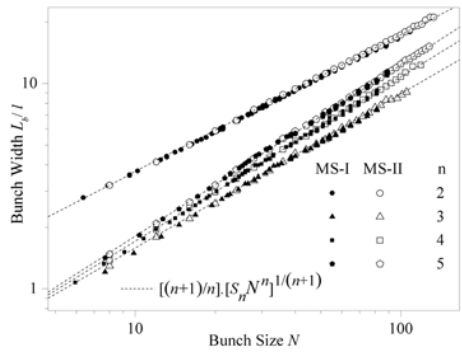

Figure 5:

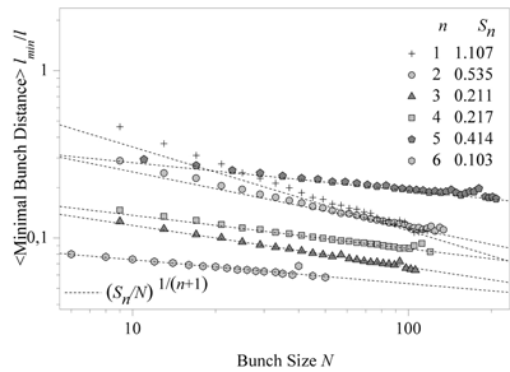

Figure 6:

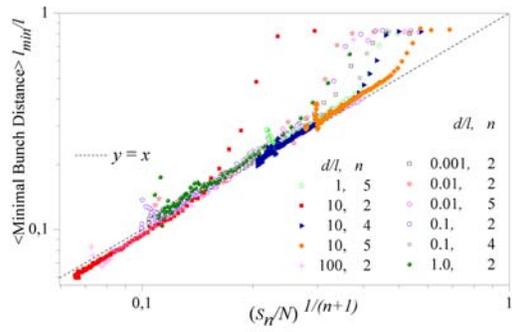

Figure 7:

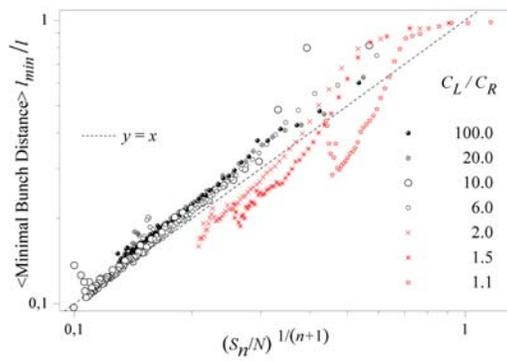

Figure 8:

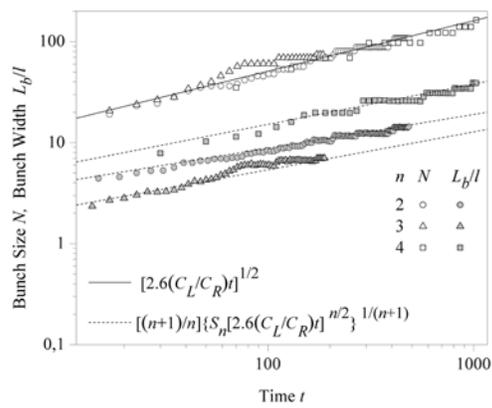

Figure 9:

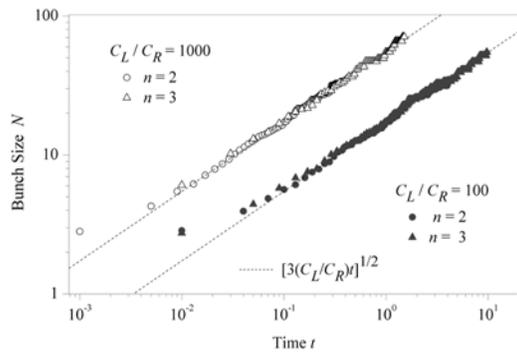